\documentclass[%
reprint,
superscriptaddress,
 amsmath,amssymb,
aps,
pra,
]{revtex4-1}

\pdfoutput=1
\usepackage{journal}
\usepackage{graphicx}
\usepackage{dcolumn}
\usepackage{bm}

\begin{document}


\title{On-chip generation and demultiplexing of quantum correlated photons using a silicon-silica monolithic photonic integration platform}

\author{Nobuyuki Matsuda}
\email{m.nobuyuki@lab.ntt.co.jp}
\affiliation{NTT Nanophotonics Center, NTT Corporation, Atsugi, Kanagawa 243-0198, Japan}
\affiliation{NTT Basic Research Laboratories, NTT Corporation, Atsugi, Kanagawa 243-0198, Japan}

\author{Peter Karkus}
\affiliation{NTT Basic Research Laboratories, NTT Corporation, Atsugi, Kanagawa 243-0198, Japan}

\author{Hidetaka Nishi}
\affiliation{NTT Nanophotonics Center, NTT Corporation, Atsugi, Kanagawa 243-0198, Japan}
\affiliation{NTT Device Technology Laboratories, NTT Corporation, Atsugi, Kanagawa 243-0198, Japan}

\author{Tai Tsuchizawa}
\affiliation{NTT Nanophotonics Center, NTT Corporation, Atsugi, Kanagawa 243-0198, Japan}
\affiliation{NTT Device Technology Laboratories, NTT Corporation, Atsugi, Kanagawa 243-0198, Japan}

\author{\\William J. Munro}
\affiliation{NTT Basic Research Laboratories, NTT Corporation, Atsugi, Kanagawa 243-0198, Japan}

\author{Hiroki Takesue}
\affiliation{NTT Basic Research Laboratories, NTT Corporation, Atsugi, Kanagawa 243-0198, Japan}

\author{Koji Yamada}
\affiliation{NTT Nanophotonics Center, NTT Corporation, Atsugi, Kanagawa 243-0198, Japan}
\affiliation{NTT Device Technology Laboratories, NTT Corporation, Atsugi, Kanagawa 243-0198, Japan}


\begin{abstract}
We demonstrate the generation and demultiplexing of quantum correlated photons on a monolithic photonic chip composed of silicon and silica-based waveguides. Photon pairs generated in a nonlinear silicon waveguide are successfully separated into two optical channels of an arrayed-waveguide grating fabricated on a silica-based waveguide platform.
\end{abstract}

\pacs{Valid PACS appear here}
\maketitle


\section{Introduction}

Integrated waveguide technology has proven useful for the large-scale integration of quantum information systems on photonic chips \cite{ladd10}, thanks to its compactness and circuit stability. In this context, intense study is under way on the development of on-chip quantum components, such as quantum processing circuits \cite{politi08,smith09,peruzzo10,bonneau12,spring13,wang14,spagnolo14}, quantum light sources \cite{sharping06,takesue07,clemmen09,xiong10,lobino11,xiong11,davanco12,horn12,matsuda12,olislager13,clark13,matsuda13,boitier14}, and single photon detectors \cite{pernice12,sahin13,najafi14}. Moreover, to fully exploit the advantages of integrated photonics, it is ideal to integrate these different components on a single substrate. Motivated by this goal, several researchers have recently demonstrated the hybrid \cite{meany14} or monolithic \cite{silverstone14,jons14,prtljaga14} integration of different quantum-optical components.

The quantum light sources include photon pair sources, which can serve as entangled photon pairs or heralded single photons. A photon pair source can be realized by employing nonlinear wave mixing in integrated waveguides such as silicon wire waveguides \cite{sharping06,takesue07,clemmen09,xiong11,davanco12,matsuda13}. Quantum circuits can also be realized by using integrated waveguides with cores made of Si \cite{bonneau12}, GaAs \cite{wang14} or silica-based materials \cite{politi08,smith09,peruzzo10,spring13,spagnolo14}. Of these approaches, silica-based waveguide technology has realized planar lightwave circuits with a significantly large scale for classical optical communication \cite{sohma04,hibino02}; this capability will facilitate the construction of large-scale quantum circuits in the near future. In addition, the low nonlinearity of silica \cite{matsuda09} helps us to avoid the generation of unwanted photons by the intense pump fields used for photon pair generation in the quantum light sources. To exploit these advantages, the integration of quantum light sources and silica waveguides is an attractive technology with which to construct on-chip quantum information systems. A significant step in this direction is the integration of a photon pair source with its interface, namely a photon-pair demultiplexer, for direct connection to a quantum circuit.

In this paper, we demonstrate the monolithic integration of a Si waveguide photon pair source and a photon-pair demutiplexer employing a silica-based arrayed waveguide grating (AWG). In Sec. 2, we investigate the photon pair generation property of the monolithic waveguide platform. By performing experiments using waveguides of various lengths, we confirm that the contribution of our silica-based waveguide to unnecessary photon generation is negligible. In Sec. 3, we demonstrate the on-chip generation and demultiplexing of a correlated photon pairs using a monolithic circuit consisting of a Si-wire photon pair source and a silica-based AWG. We show that the chip is capable of generating quantum correlated photons and guiding them into different output ports

\section{Photon pair generation properties on a silicon-silica monolithic waveguide platform}

\begin{figure}
	\centering\includegraphics[width=8.6cm]{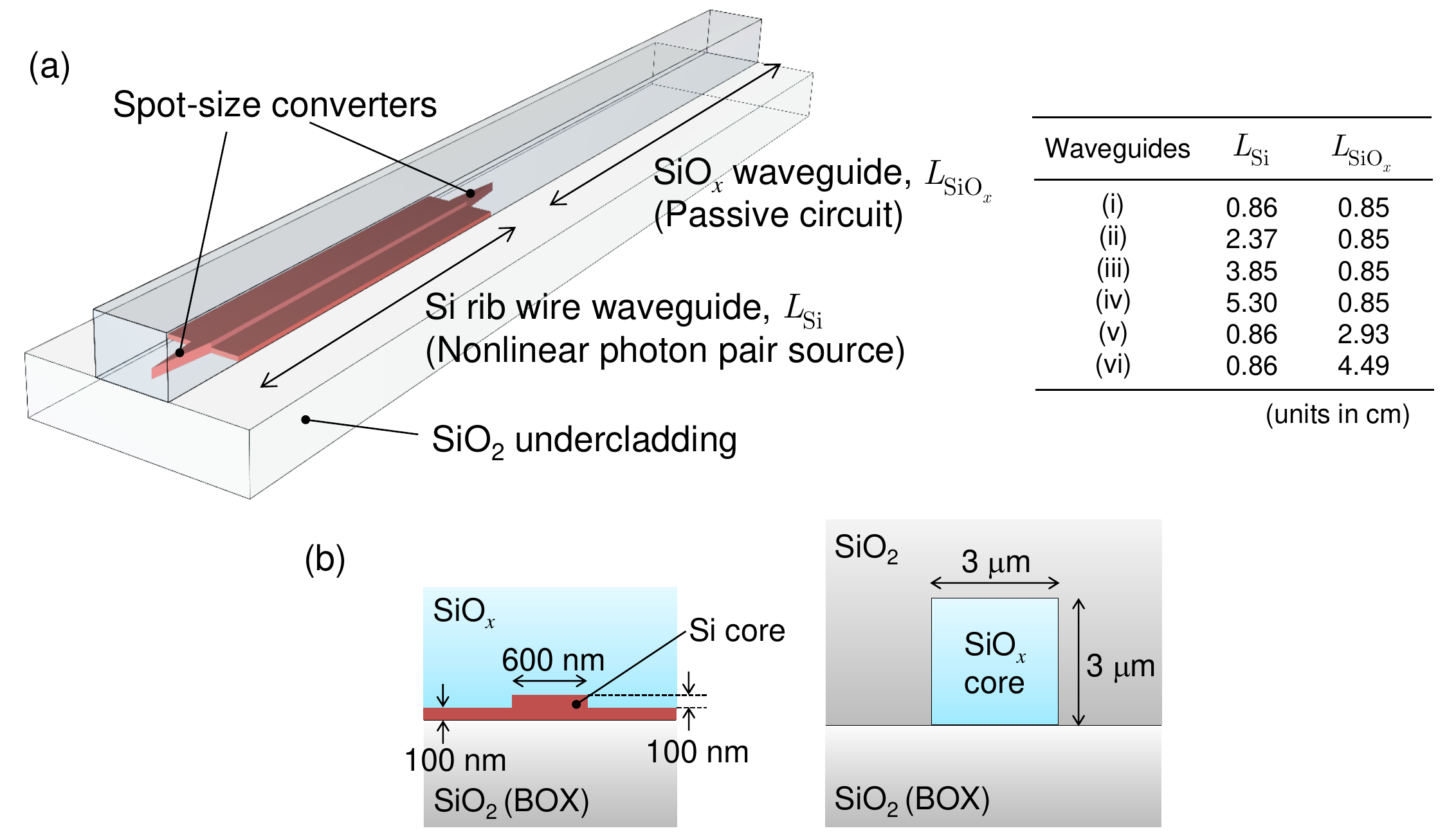}
	\caption{(a) Si-silica monolithic waveguide platform for an integrated photonic quantum information system. $L_{\rm Si}$ and $L_{\rm{SiO}_x}$ are the lengths of each waveguide section. The table shows the waveguide length combinations used in the experiment. (b) Cross-sectional views of each waveguide section. The silica overcladding and Si substrate are omitted for clarity.}
	\label{fig:hybrid}
\end{figure}

Figure \ref{fig:hybrid}(a) is a schematic diagram of a monolithic waveguide platform made of Si and silicon-rich silica (SiO$_{x}$) \cite{nishi10}. We first fabricate a silicon wire rib waveguide by electron-beam lithography and electron-cyclotron resonance (ECR) plasma etching on a silicon-on-insulator (SOI) substrate. The SiO$_{x}$ waveguides are fabricated in a region from which the top Si layer of SOI substrate has been removed by reactive ion etching (RIE). In that region, the SiO$_{x}$ layer is deposited by ECR plasma-enhanced chemical vapor deposition (PE-CVD) at low temperature. Then the cores of the SiO$_{x}$ waveguides are fabricated by photolithography and RIE. Finally, the SiO$_{2}$ layer is deposited by ECR PE-CVD. The SiO$_{x}$ waveguides have a core-cladding index contrast of $\sim$ 3\%. We also fabricate spot-size converters (SSCs) with a tapered Si waveguide for the low-loss connection of the Si and SiO$_{x}$ waveguides. The cross-sectional dimensions of each waveguide are shown in Fig. \ref{fig:hybrid}(b). We use several devices with different combinations of Si and SiO$_{x}$ waveguide lengths, $L_{\rm{Si}}$ and $L_{\rm{SiO}_x}$, as shown in the table in Fig. \ref{fig:hybrid}(a). The propagation losses of the Si and SiO$_{x}$ waveguides, $\alpha_{\rm{Si}}$ and $\alpha_{\rm{SiO}_x}$, are estimated by the cut-back method to be approximately -- 2.1 and -- 1.8 dB/cm for the fundamental transverse-electric (TE) mode, respectively.

\begin{figure}
	\centering\includegraphics[width=8.6cm]{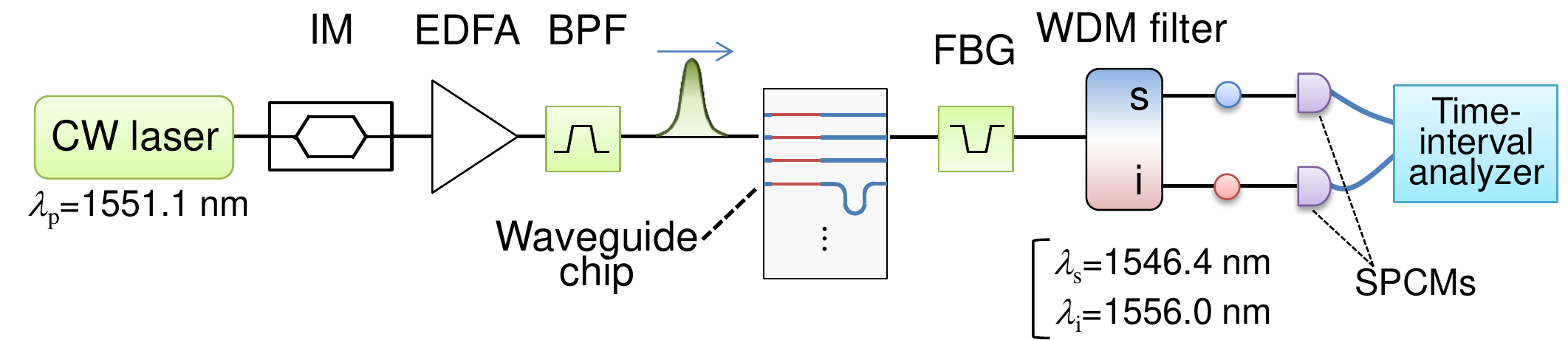}
	\caption{Experimental setup for photon-pair generation experiment of Si-silica monolithic waveguide chip. IM: LiNbO$_{3}$ intensity modulator, EDFA: erbium-doped fiber amplifier, BPF: band-pass filter, WDM filter: wavelength-division multiplexing filter, SPCM: single-photon counting module.}
	\label{fig:setup1}
\end{figure}

To investigate the nonlinearity in the monolithic waveguide, we undertake a photon pair generation experiment using the device with a cascaded structure consisting of a Si and a SiO$_{x}$ waveguide shown in Fig. \ref{fig:hybrid}(a). The experimental setup is depicted in Fig. \ref{fig:setup1}. The LiNbO$_{3}$ intensity modulator (IM) modulates a continuous beam from the light source operating at a wavelength $\lambda_{\rm{p}}$ of 1551.1 nm into a train of pump pulses with a temporal full-width at half maximum (FWHM) $\Delta t$ of 200 ps and a repetition rate $R$ of 100 MHz. The pulses are amplified by an erbium-doped fiber amplifier (EDFA), filtered with a band-pass filter BPF (3-dB bandwidth: 0.2 nm) to eliminate amplified spontaneous emission noise, and then launched into the waveguides with a lensed fiber from the Si waveguide side. The input polarization is set at the fundamental TE mode. The in- and out-coupling efficiency with the chip, $\eta_{\rm{couple}}$, is approximately -- 1 dB/facet.

In nonlinear waveguides such as a Si waveguide, a correlated pair of signal and idler photons are created via a $\chi^{(3)}$ spontaneous four-wave mixing (SFWM) process following the annihilation of two photons inside the pump pulse \cite{sharping06,takesue07}. The $\chi^{(3)}$ nonlinearity of Si is 200 times higher than that of silica. In addition, the core area of a Si waveguide is approximately two orders of magnitude smaller than that of our SiO$_{x}$ waveguide. Thus we can expect the Si part to play a major role in correlated photon pair generation. However, this should be confirmed experimentally since SiO$_{x}$ has a material composition different from that of standard silica. We should also investigate the noise photon generation characteristics in the SiO$_{x}$ waveguide, since the Raman scattered photons in fused silica waveguides such as optical fibers \cite{fiorentino02,takesue04} are a potential source of noise.

The optical fields output from the side of the SiO$_{x}$ waveguide including the correlated photons were collected by another lensed fiber. Then, the light was introduced into the fiber-Bragg grating (FBG) filter and the wavelength-division multiplexing (WDM) filter, which separated the signal and idler photons into different fiber channels. Here the total pump-wavelength suppression of the FBG and WDM filters exceeds 130 dB. Each output port of the WDM filter has center wavelengths of 1546.4 nm ($\lambda_{\rm{s}}$) and 1556.0 nm ($\lambda_{\rm{i}}$) with a passband width $\Delta \nu$ of 0.12 THz (0.96 nm). Finally, the photons are received by avalanche-photodiode-based single photon counting modules (SPCMs) (id210, id Quantique) that operated at a gate frequency of 100 MHz synchronized with the pump repetition rate $R$. The quantum efficiency $\eta_{\rm{QE}}$, gate width, dark count rate $d$, and dead time of the detectors were 21 \%, 1.0 ns, 2.1 kHz, and 10 $\mu$s, respectively. The overall transmittance of the filtration system $\eta_{\rm{f}}$ is approximately -- 3.8 dB. The raw coincidence rate $D_{\rm{c}}$ (including the accidental coincidence count) and the raw accidental coincidence rate $D_{\rm{c,a}}$ were determined by measuring the time correlation of the signals output from the two SPCMs using a time-interval analyzer.

\begin{figure}
	\centering\includegraphics[width=8.6cm]{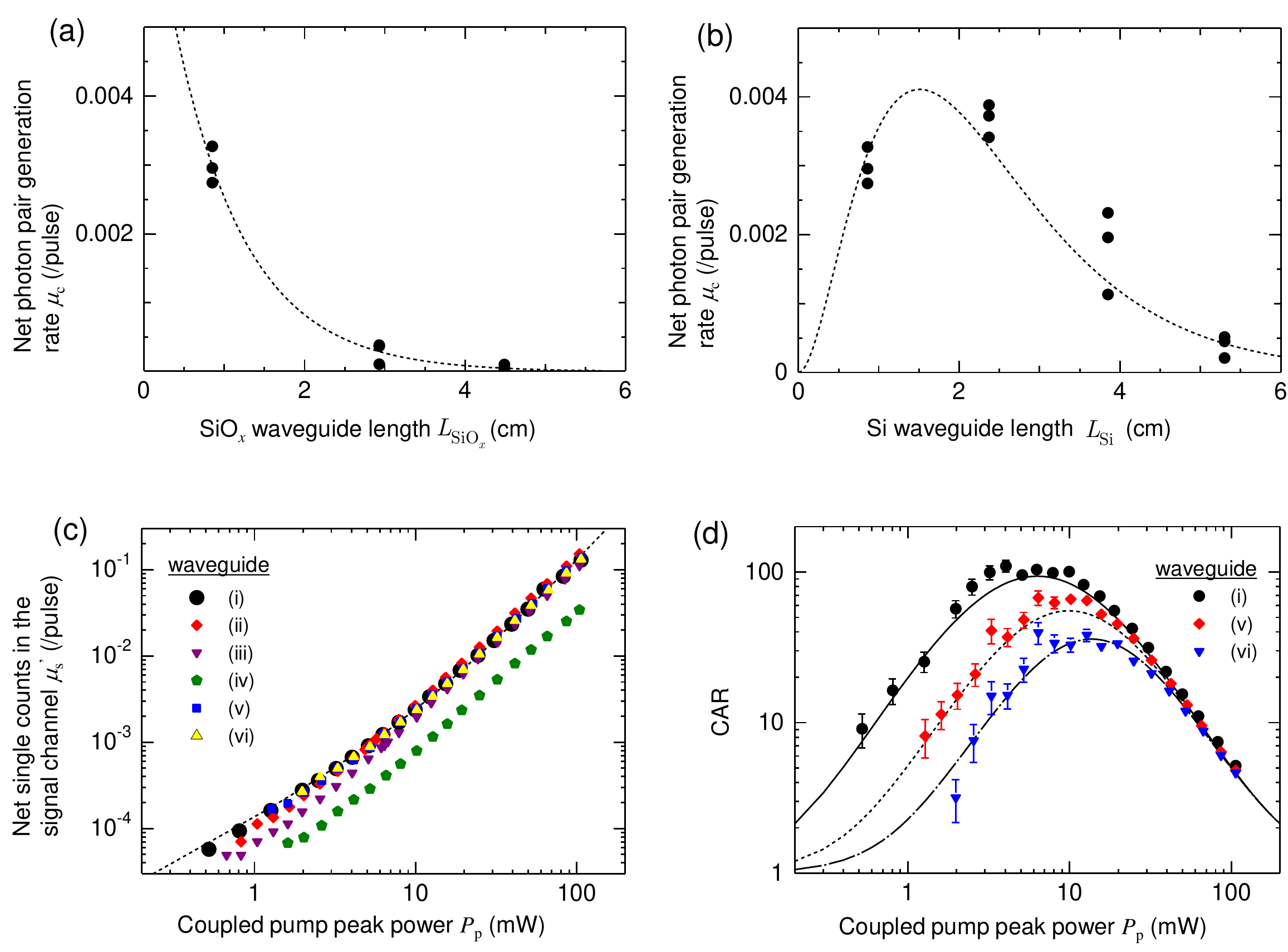}
	\caption{Photon pair generation rate as a function of (a) SiO$_{x}$ and (b) Si length, respectively. The pump peak power $P_{\rm p}$ is fixed at 37 mW. The dashed curves show the fitting results as described in the text. (c) The net photon generation rate inside the wavelength band for the signal photons at the end of the Si waveguide versus $P_{\rm p}$. (d) Coincidence to accidental coincidence ratio (CAR) values as a function of $P_{\rm p}$. The curves show the results of numerical calculations as described in the text.} 
	\label{fig:result1}
\end{figure}

Figures \ref{fig:result1}(a) and 3(b) are the net photon pair generation rate (at the waveguide output end of the SiO$_{x}$ waveguide side) as a function of the waveguide length. Fig. \ref{fig:result1}(a) shows the $L_{\rm{SiO}_x}$ dependence obtained using waveguides (i), (v), (vi), whereas Fig. \ref{fig:result1}(b) is the $L_{\rm{Si}}$ dependence with waveguides (i) to (iv). We used three waveguides for each length condition. Here we estimated the net photon pair generation rate at the end of the SiO$_{x}$ waveguide, $\mu_{\rm{c}}$, using
\begin{equation}
	\mu_{\rm{c}} = \frac{D_{\rm{c}} - D_{\rm{c,a}}}{R \eta_{\rm total}^2},
	\label{eq:pairest}
\end{equation}
where $\eta_{\rm total} = \eta_{\rm{couple}} \eta_{\rm{f}} \eta_{\rm{QE}} \eta_{\rm{gate}} $ with $\eta_{\rm{gate}}$ being the ratio of the active gates to the 100 MHz clock rate \cite{matsuda13}. The number of active detector gates decreases due to the finite detector dead time set in our experiment. The measurement time was 60 s for each data point to obtain good statistics.

In more detail, Fig. \ref{fig:result1}(a) shows that the pair generation rate decreases monotonically with increases in $L_{\rm{SiO}_x}$, in contrast to the $L_{\rm{Si}}$ dependence shown in Fig. \ref{fig:result1}(b). This strongly suggests that the contribution of the SiO$_{x}$ waveguide as a photon pair source is negligible. Assuming that the SiO$_{x}$ waveguide is a passive transmission line to the photon pairs generated in the Si waveguide section, $\mu_{\rm{c}}$ should follow $\mu_{\rm{c}} \propto \eta_{\rm SiO_{\it x}}^2$, where $\eta_{\rm SiO_{\it x}} = \mathrm{e}^{- \alpha_{\rm SiO_{\it x}} L_{\rm SiO_{\it x}} } $ is the transmittance in the SiO$_{x}$ waveguide. A fitted function with this relation is shown as a dashed curve in Fig. \ref{fig:result1}(b). The experimental results are well explained by the fitting. From the fitting we obtained an $\alpha_{\rm SiO_{\it x}}$ value of -- 2.4 dB/cm, which is similar to the value obtained with the cut-back method. These results suggest that the SiO$_{x}$ waveguide works as a passive circuit without the creation of a significant number of unwanted photon pairs inside it.

On the other hand, in Fig. \ref{fig:result1}(b) we find that $\mu_{\rm c}$ increases with increases in $L_{\rm Si}$, and then starts to decrease in the $L_{\rm Si} >$ 3 cm region. The latter is considered to be due to the propagation loss in the Si waveguide. Taking the linear propagation loss into account, the photon pair generation rate at the end of the first Si waveguide $\mu_{\rm c}'$ can be written as
\begin{equation}
	\mu_{\rm c}' = \Delta \nu \Delta t ( \gamma P_{\rm p} L_{\rm eff} )^2 \eta_{\rm Si}^2,
	\label{eq:pairrate}
\end{equation}
where $\gamma$ is the nonlinear constant of the Si waveguide, $P_{\rm p} = P / (R \Delta t)$ is the coupled pump peak power, and $L_{\rm eff}$ is the effective Si waveguide length associated with $L_{\rm eff} = \frac{1-\mathrm{e}^{- \alpha_{\rm Si} L_{\rm Si}}}{\alpha}$. $\eta_{\rm Si} = \mathrm{e}^{- \alpha_{\rm Si} L_{\rm Si}}$ is the linear transmittance of the Si waveguide that causes the intrinsic loss of the photon pairs \cite{husko13, lin06}. Here we assumed that $\alpha_{\rm Si}$ has no wavelength dependence.
The dashed curve in Fig. \ref{fig:result1}(b) shows a fitted function using $\mu_{\rm c} = \eta_{\rm SiO_{\it x}}^2 \mu_{\rm c}'$ and Eq. (\ref{eq:pairrate}) with a fitting parameters $\gamma$ = 161 /W/m and $\alpha_{\it Si}$ = 2.0 dB/cm. The experimental data are well explained by the fitting. The $\gamma$ value obtained from the fitting is slightly lower than that of channel-type Si wire waveguides \cite{takesue07, clemmen09}. This is because the present rib-type waveguide has a larger effective mode area than a channel waveguide.

Noise photons, such as Raman scattered photons, can contaminate wavelength channels for the correlated photons. To investigate the generation of noise photons in our SiO$_{x}$ waveguide, we plot single count rate $\mu_{\rm s}'$ as a function of $P_{\rm p}$ in Fig. \ref{fig:result1}(c). Here $\mu_{\rm s}'$ is the photon generation rate in the signal wavelength channel at the end of the Si waveguide part (before the SiO$_{x}$ waveguide) estimated by $\mu_{\rm s}' = \frac{N_{\rm s}}{R \eta_{\rm{total}} \eta_{\rm SiO_{\it x}}}$, where $N_{\rm s}$ is the raw single count rate measured by the SPCM set in the signal channel excluding the dark count rate of the detector. From Fig. \ref{fig:result1}(c), we find that the $\mu_{\rm s}'$ values remain the same regardless of $L_{\rm SiO_{\it x}}$ (by comparing the results obtained with waveguides (i), (v) and (vi)). This means that a negligible number of noise photons are generated in the SiO$_{x}$ waveguide. The dashed curve shows a second-order polynomial fitting to the data obtained for the waveguide (i). In addition to the $P_{\rm p}^2$ component that originated from the SFWM, we can see the $P_{\rm p}^1$ component in the low excitation regime. This indicates that processes other than SFWM, for example inelastic scattering in the Si waveguide part, contribute little \cite{clemmen12}.

Finally, Fig. \ref{fig:result1}(d) shows the measured coincidence to accidental coincidence ratio (CAR = $D_{\rm c}/D_{c,a}$) from three waveguides with the same $L_{\rm Si}$ with respect to $P_{\rm p}$. The maximum CAR of around 100 shows the strong quantum correlation of the photon pairs. We see that the overall CAR values decrease with increases in $L_{\rm SiO_{\it x}}$. If the SiO$_{x}$ waveguides generated a negligible number of noise photons, this reduction should be explained by the linear propagation loss of the photons in the SiO$_{x}$ waveguides. To confirm this, we estimated CAR using
\begin{equation}
	\rm{CAR} = \frac{\eta^2 \eta_{\rm SiO_{\it x}}^2 \mu_{\rm c} }{ (\eta_{\rm total} \mu_{\rm s}' \it + d ) (\eta_{\rm total} \mu_{\rm i}' \it + d )} + 1,
	\label{eq:car}
\end{equation}
here we calculated $\mu_{\it c}$ using Eq. (2) with $\gamma$ = 161 /W/m, which we obtained from the fitting above. For $\mu_{\rm s}'$ we used the fitted functions for waveguide (i) represented by the dashed curve in Fig. \ref{fig:result1}(c); we obtained the $\mu_{\rm i}'$ function in the same way. The estimated CAR is shown by the solid curve in Fig. \ref{fig:setup1}(d), which agrees well with the experimental data for waveguide (i). Next, we replace the $L_{\rm SiO_{\it x}}$ value with 2.93 and 4.49 cm in the calculation above, and plot the results as dashed and dot-dashed curves, respectively. The two curves well describe the experimental data obtained with waveguides (v) and (vi). Thus, we confirmed that the decrease in CAR with increases in SiO$_{x}$ waveguide length can be explained by the photon loss in the SiO$_{x}$ part. This also indicates that no significant noise photons are created in the SiO$_{x}$ waveguide.

\section{On-chip generation and wavelength-division demultiplexing of photon pairs}

We have shown that our SiO$_{x}$ waveguide works as a low-nonlinear circuit without creating a significant number of noise photons. Next we attempt to monolithically integrate the Si-wire photon pair source and its passive wavelength demultiplexing filter on the same chip.

\begin{figure}
	\centering\includegraphics[width=8.6cm]{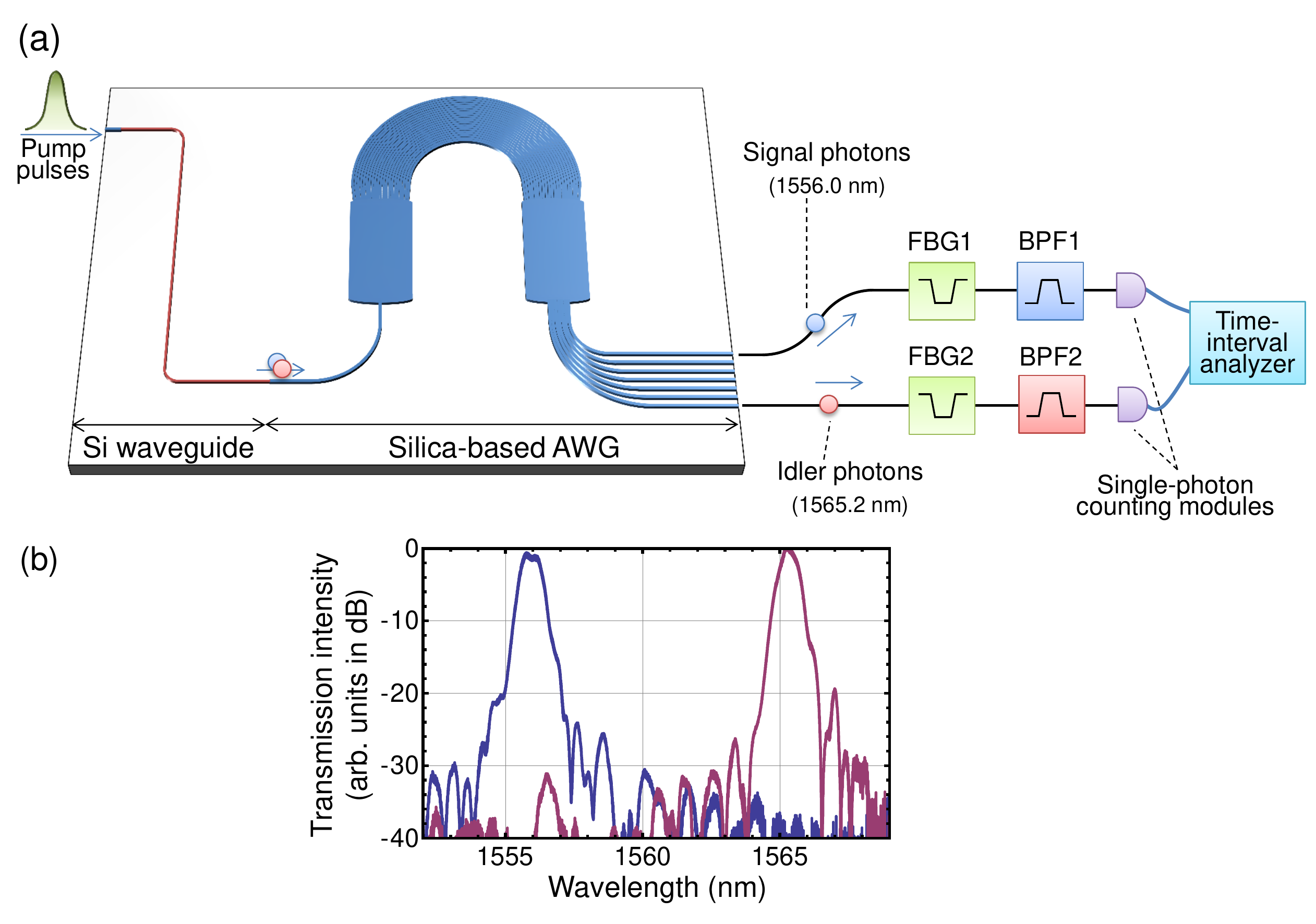}
	\caption{(a) Monolithic chip housing a Si wire photon pair source and a silica-based arrayed waveguide grating, illustrated with the experimental setup. (b) Transmission spectra of AWG output ports, from which photon pairs are collected.}
	\label{fig:awg}
\end{figure}

Figure \ref{fig:awg}(a) is a schematic diagram of the monolithic device fabricated in the manner described in Sec. 2, together with the experimental setup. In the device, correlated photon pairs are created via the SFWM in the first Si rib waveguide ($L_{\it Si}$ = 1.37 cm), and subsequently spectrally separated by the on-chip SiO$_{x}$ AWG into different output channels. Our AWG has 16 output channels designed to have a 200 GHz channel spacing. AWGs are commonly used for separating photon pairs generated via SFWM in integrated waveguides \cite{takesue07, xiong10, lobino11, xiong11, clark13, matsuda13}. The monolithic integration of the AWG and SFWM-based photon pair source thus provides a compact and stable photon pair source and a heralded single photon source. It is also a key technology for the implementation of a multiplexed single photon source \cite{meany14}. Moreover, our monolithic AWG can be used as an interface between a photon pair source and a linear-optic quantum circuit.

The pump pulses for the experiment are obtained using the setup shown in Fig. \ref{fig:setup1}. We choose to collect photons output from a pair of waveguides that are 3 channels away from the center output port. With this channel separation we obtain a pump-to-signal (or idler) wavelength separation similar to that of the experiments described in Sec. 2.  We show the transmission spectra of the two AWG outputs in Fig. \ref{fig:awg}(b). Here the 3-dB passband widths of the transmission windows are approximately 80 GHz. The SSCs with Si tapers are fabricated between the Si waveguide and the AWG for their low-loss connection. The output optical fields are collected by optical fibers with a high numerical aperture. Then the photons are introduced into spectral filters, each of which consists of an FBG notch filter and a BPF for the suppression of residual pump fields. The 3-dB bandwidth of the BPFs $\Delta \nu$ are 100 GHz, which covers the AWG passbands. Finally, the photons are received by SPCMs and a coincidence measurement is performed with a time-interval analyzer. The overall transmittance of the filters $\eta_{\rm f}$ is --2.8 dB and the AWG insertion loss $\eta_{\rm AWG}$ is --7.7 dB. The quantum efficiency $\eta_{\rm QE}$ and dark count rate $d$ of the SPCM are 24\% and 5.1 kHz, respectively, in the present experiment.

\begin{figure}
	\centering\includegraphics[width=8.6cm]{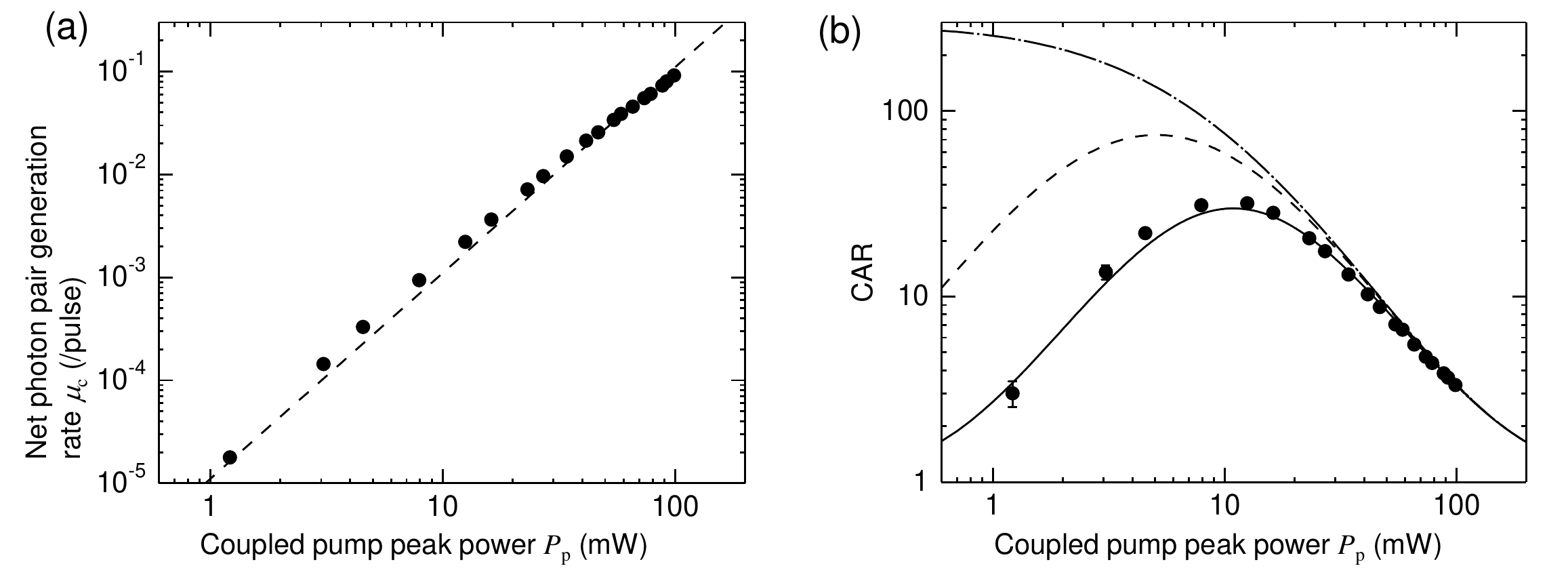}
	\caption{(a) Net photon pair generation rate as a function of pump peak power $P_{\rm p}$. The dashed curve shows the theoretical calculation result. (b) CAR values as a function of $P_{\rm p}$. The solid curve shows a calculation result. The dashed and dot-dashed curves are estimations of the lower AWG insertion loss and lower dark count rate, respectively, described in the text.}
	\label{fig:result2}
\end{figure}

To complete the integration and characterization of the device, Fig. \ref{fig:result2}(a) shows the net photon pair generation rate estimated using Eq. (\ref{eq:pairest}) as a function of the pump peak power $P_{\rm p}$. The data exhibit good $P_{\rm p}^2$ dependence, indicating photon pair generation via the SFWM process. The solid line shows the estimation obtained with Eq. (\ref{eq:pairrate}) and $\mu_{\rm c} = \eta_{\rm AWG}^2 \mu_{\rm c}'$ using the same $\gamma$ value of 161 /W/m obtained from the fitting in Sec. 2. The experimental result agrees well with the calculation. Hence the observed photon pairs are considered to be generated via SFWM in the Si waveguide.

Figure \ref{fig:result2}(b) shows the measured CAR as a function of $P_{\rm p}$. We obtained a maximum CAR of around 30. The value is well above CAR = 2, which is the limit accessible with classical light \cite{loudon}. This indicates that the quantum correlation between photons was preserved even after they had passed through the integrated AWG. Thus, our chip successfully generated and demultiplexed quantum correlated photons on the monolithic device. Next we analyze the obtained CAR values. The solid curve shows CAR values estimated using Eq. (\ref{eq:car}) in accordance with the procedure employed to obtain the solid curve in Fig. \ref{fig:result1}(c). The curve agrees well with the experimental data. The dashed curve shows the CAR values for $\eta_{\rm AWG}$ = 0, exhibiting a maximum CAR of up to 80. Hence, reducing the AWG insertion loss is effective in improving CAR. We also show the CAR values obtained when the detector dark count rate $d$ = 20 (Hz). This suggests that we can further improve the maximum CAR by using SPCMs with low dark count rates such as superconducting single photon detectors with similar dark count rates \cite{inagaki13}.

\section{Conclusion}
We have demonstrated the on-chip generation and demultiplexing of quantum correlated photon pairs using a monolithic waveguide platform composed of Si and silica-based waveguides. Furthermore, we have shown that the silica part of the monolithic platform does not contribute to noise photon generation. The device can be used as a compact correlated photon pair source, and will be useful for many quantum information applications including wavelength-division multiplexing quantum communication technologies \cite{yoshino12} and heralded single photon sources \cite{davanco12}. Moreover, the silica-based AWG can provide an interface between a Si-based photon pair source and silica-based lightwave circuits, which are useful as linear-optics-based quantum circuits. The wavelength-multiplexing capability is also beneficial for constructing a circuit that harnesses high-dimensional quantum states using path and frequency degrees of freedom \cite{schaeff12}. Thus, the present platform will prove useful for monolithic source-circuit integration with a view to achieving the full-scale integration of on-chip quantum processors.

\section*{Acknowledgement}
We are grateful to Dr. Kaoru Shimizu for fruitful discussions.


\end{document}